\documentstyle[twoside,fleqn,espcrc2]{article}
\title{Signal for CP violation in $B^{\pm} \to P {\bar P} \pi^{\pm}$ decays}
\author{B. Bajc\address{J. Stefan Institute, Jamova 39, P. O. Box 3000, 1001 
Ljubljana, 
Slovenia}, S. Fajfer$^{\rm a}$, R. J. Oakes\address{Department of Physics and 
Astronomy, Northwestern University, 
Evanston, Il 60208, U.S.A.},  
T.N. Pham\address{Centre de Physique Theorique, Centre National de la Recherche 
Scientifique, 
UPR A0014,\\ Ecole Polytechnique, 91128 Palaiseau Cedex, France} and S. Prelov\v 
sek$^{\rm a}$}
\begin{document}

\begin{abstract}
We analyze  the partial rate asymmetry  in  
$B^{\pm} \to P {\bar P} \pi^{\pm}$  
decays ($ P = \pi^+, K ^+ , \pi^0, \eta$) which results from the 
interference of 
the nonresonant decay amplitude and the resonant amplitude for 
$B^{\pm} \to \chi_{c0} \pi^{\pm} $
followed by the decay $\chi_{c0} \to P {\bar P} $.  The CP violating phase 
$\gamma$ can be extracted from the measured asymmetry.  
We find that the partial rate asymmetry for $B^\pm \to \pi^+ \pi^- \pi^\pm$ is 
$0.33~sin \gamma$, while for $B^\pm \to K^+ K^-\pi^\pm$ it amounts  
$0.45~ sin \gamma$. 
\end{abstract}
\maketitle

The measurement  of CP asymmetries in charged B  meson decays  might provide us 
 with a first demonstration 
of CP violation outside the K system \cite{Nir,Ali}. 
One possibility to measure the CP odd phase $\gamma = arg (V_{ub}^*)$ has been 
 suggested  in \cite{DEHT}. 
 In this work the asymmetry appears as   
a result of the interference of the nonresonant decay 
$B^{\pm}\to P\bar  P \pi^{\pm}$ amplitude  
and the resonant  $B^{\pm} \to \chi_{c0} \pi^{\pm} $ $\to P\bar  P \pi^{\pm}$, 
where $\chi_{c0}(3.4)$ is a $c {\bar c} $ scalar \cite{DEHT,EGM}. 
The absorptive phase necessary to observe CP violation in 
partial rate asymmetries is provided by the $\chi_{c0}$ width.

On the experimental side the CLEO collaboration has reported upper limits 
on $B ^+ \to h^+ h^+ h^-$ 
nonresonant decays \cite{CLEO}. They found that the upper limit on the 
branching ratio $BR(B^+ \to \pi^+ \pi^- \pi^+) \le 4.1 \times 10^{-5}$ and 
$BR(B^+ \to K^+ K^-\pi^+) \le 7.5 \times 10^{-5}$. 
The authors of \cite{DEHT} estimated theoretically that 
the branching ratio for $B^+ \to \pi^+ \pi^- \pi^+$ ranges from 
 $1.5 \times10^{-5}$  to $ 8.4 \times10^{-5}$.
  
Motivated by the theoretical expectation and the 
experimental result,  
we investigate the $B^{\pm} \to P {\bar P} \pi^{\pm}$ nonresonant decay 
widths, where $ P = \pi^+, K ^+ , \pi^0, \eta$. 
We also study   the resonant decay amplitudes of 
$B^{\pm} \to P {\bar P} \pi^{\pm}$ 
arising from 
$\chi_{c0}$  decays into $P {\bar P}$, $ P = \pi^+, K ^+ , \pi^0, \eta$. 
The branching ratios of the $\chi_{c0}$ decays into these 
modes have been reported \cite{PDG}.  
 We point out that  we are interested only in the kinematical region 
 where the $P\bar P$ invariant mass is close to the $\chi_{c0}$ mass, 
 like in \cite{DEHT}. Thus $P {\bar P}$  arising from 
resonances such as $\rho$, and other possible resonances 
will  not be considered in this paper.   
In the analysis of the nonresonant decay amplitude of 
$B^{\pm} \to P {\bar P} \pi^{\pm}$ 
we use the factorization approximation, which shows that 
the main contribution comes from  the product 
$< P \bar P| ({\bar u} b)_{V - A}| B^->$ $ < \pi^- | ({\bar d} u)_{V -A} |0>$   
($(\bar q_1 q_2 )_{V-A}$ 
denotes  $\bar q_1 \gamma_{\mu} (1- \gamma_5)q_2$ ). 
For the calculation of the matrix element 
$< P \bar P| ({\bar u} b)_{V - A}| B^->$ 
we use the results obtained in \cite{BFOP}, where the nonresonant 
$D^+ \to K^- \pi^+ l \nu$ decay 
was analyzed. The experimental result for the branching ratio of 
the nonresonant $D^+ \to K^- \pi^+ l \nu$ decay was successfully reproduced   
within a framework (hybrid model) which combines the 
heavy quark effective theory (HQET) 
and the chiral Lagrangian (CHPT). The combination  of heavy quark symmetry and  
chiral symmetry was quite 
successful in the 
analysis of D meson semileptonic decays \cite{caspr,BFO},  and references 
therein.
The heavy quark symmetry is expected to work better 
for B mesons \cite{caspr}. However, 
new difficulties might appear in B decays, due to the large 
energies of light mesons in 
the final state. It is known, however,  that the combination of 
HQET and CHPT is valid at 
small 
recoil momentum. The details of 
our approach developed for D meson semileptonic decays, 
can be found in \cite{BFOP,BFO}. We systematically apply the 
hybrid model  in the calculation of the 
$B^{\pm} \to P {\bar P} \pi^{\pm}$ amplitude and we find that 
in a weak transition of $B \to P \bar P$ new important contributions 
arise, which were not taken into account  in \cite{DEHT}.\\

The weak effective Lagrangian for the nonleptonic Cabibbo 
suppressed $B$ meson decays is given by \cite{DEHT} 
\begin{equation}
{\cal L}_{w} = - \frac{G_F}{{\sqrt 2}} V_{ud}^* V_{ub} (a_1^{eff} O_1 + 
a_2^{eff} O_2)
\label{eq1}
\end{equation}
where $a_1^{eff}  \simeq 1.08$ and $a_2^{eff} \simeq 0.21$  \cite{Ali}, 
$O_1 = ({\bar u} b)_{V-A} ({\bar d} u)_{V-A}$ and 
$O_2 = ({\bar u} u)_{V-A} ({\bar d} b)_{V-A}$. 
The quark current required in the weak Hamiltonian (\ref{eq1}) 
is expressed  
in terms of meson fields \cite{BFOP}. 
In our calculations we will follow the approach 
described in \cite{BFOP} and \cite{BFO}: we use the Feynman rules for the 
vertices near and outside the zero-recoil region, {\it but we include the 
complete 
propagators instead of using the HQET propagator} \cite{caspr}. \\

Using the factorization approach \cite{DEHT},  we analyze all 
possible contributions to the $B^{\pm} \to \pi^{\pm} P{\bar P}$ 
nonresonant  amplitude.  
We notice  that the  
main contribution to the amplitude for nonresonat 
$B^\pm \to P \bar P \pi^\pm$ decay 
comes from the matrix element of 
$< P(p_1) \bar P(p_2) |$ $({\bar u} b)_{V -A} | B^- (p_B)>$. 
Following the analysis described in \cite{BFOP}, 
we write 
\begin{eqnarray}
\label{wwh}
< \pi^- (p_1) \pi^+ (p_2) | {\bar u} \gamma_{\mu} (1 - \gamma_5) b | B^- (p_B)>
&\!\!\! = \!\!\!&\nonumber\\
ir(p_B-p_2-p_1)_\mu
+iw_+(p_2+p_1)_\mu &+&\nonumber\\ iw_-(p_2-p_1)_\mu + 
2h\epsilon_{\mu\alpha\beta\gamma}p_B^\alpha p_2^\beta p_1^\gamma\;.
\end{eqnarray}
In our case only the nonresonant form factors $w_-^{nr}$ and 
$w_+^{nr}$ contribute. The contribution proportional to $r$ is of   
order $m_{\pi}^2$ and  therefore we can safely neglect it. 
In the case of $B^- \to \pi^-\pi^- \pi^+$ decay the form factors $w_+$ 
and $w_-$ are given by 
\begin{eqnarray}
w_+^{nr}(s,t)  =  - \frac{g}{f_{\pi}^2} 
\frac{f_{B*} m_{B*}^{3/2} m_B^{1/2}}{s - m_{B*}^2}& \times& \nonumber\\
\{1 - \frac{1}{2 m_{B*}^2}(m_B^2 -m_{\pi}^2 -s)\}
+  \frac{f_B}{ 2 f_{\pi}^2} & - &\nonumber\\ 
 \frac{{\sqrt m_B} \alpha_2}{ 2 f_{\pi}^2} 
\frac{1}{m_B^2}(2 s + t - m_B^2 - 3 m_{\pi}^2 ), 
\label{w+1}
\end{eqnarray}
\begin{eqnarray} 
w_-^{nr}(s)  =   \frac{g}{f_{\pi}^2} 
\frac{f_{B*} m_{B*}^{3/2} m_B^{1/2}}{s - m_{B*}^2}& \times&\nonumber\\ 
\{1   +  \frac{1}{2 m_{B*}^2}(m_B^2 -m_{\pi}^2 -s)\}
 +\frac{{\sqrt m_B} \alpha_1}{  f_{\pi}^2}.    
\label{w-1}
\end{eqnarray}
In the above expressions $f_{B,(B*)}$  are B meson decay constants, 
the parameters $\alpha_{1,2}$ are fixed in \cite{BFO} using the data from
 semileptonic 
$D^0 \to \bar K^{*0} l \nu_l$ decay. The $s$, $t$ and $u$ are defined as 
usual: $s =(p_B - p_3)^2 = (p_2+ p_1)^2$, $t =(p_B- p_1)^2 = (p_1 + p_2)^2$ 
 and $u = (p_B - p_2)^2 = (p_1+ p_3)^2$. 
The decay constants are related by the heavy quark symmetry \cite{caspr}. 
We use  $f_D \simeq 200$ $ MeV$ \cite{BFO}, giving $f_B \simeq 120$ $ MeV$. 
Then,  we use $|V_{ub}| = 0.003$ \cite{PDG}.  
The parameters $\alpha_1$ and $\alpha_2$ 
must be specified for $B$ meson decays. 
It is easy to find that the soft scaling \cite{caspr} 
of the axial form factors $A_1$ and $A_2$, 
leads to the following relations: 
$\alpha_1^{DK*} = \alpha_1^{B\rho}$, while $\alpha_2^{DK*}/m_D = 
\alpha_2^{B\rho} /m_B$.
We take the parameters $\alpha_{1,2}^{DK*}$ from  
$D$ meson semileptonic decays \cite{BFOP,BFO}.  
branching ratio $B ^-\to \pi^- \pi^+ \pi^-$

The parameter $g = 0.3 \pm 0.1$ is determined  from 
$D^* \to D \pi$ decays \cite{PC}. 
>From $D^0 \to K^- l^+ \nu$ we have found $g = 0.15 \pm 0.08$ \cite{BFOsasa}.  
In the present calculations we  take the range $0.2 \leq g \leq 0.23$ and 
we select 
$\alpha_1^{DK*} = -0.13$ $GeV^{1/2}$, $\alpha_2^{DK*} = -0.13$ $GeV^{1/2}$,  
 giving  $3.4 \times 10^{-5}\leq$ $BR(B^- \to \pi^- \pi^+ \pi^+) \leq$ 
 $ 3.8 \times 10^{-5}$.  
All other combinations of parameters give the branching ratio too 
large in comparison with the experimental limit. 
Using the same set of parameters, we find 
$1.4 \times 10^{-5}\leq$$BR(B^-\to K^- K^+ \pi^-) \leq$ $1.5 \times 10^{-5}$. 
We calculate the following limits for 
the branching ratios $4.6 \times 10^{-6}\leq$ 
$BR(B^- \to \pi^- \pi^0 \pi^0)\leq$ $ 6.1\times 10^{-6}$ 
and $6.4 \times 10^{-7}\leq$ 
$BR(B^- \to \pi^- \eta \eta) \leq 8.5 \times 10^{-7}$. 
We find  that the contributions
proportional to the $\alpha_{1,2}$ in 
the calculated branching ratio are very important. \\

The resonant decay amplitude for $B^- \to  \chi_{0c} \pi^- \to \pi^+ \pi^- 
\pi^-$
is determined by the narrow width approximation 
\begin{eqnarray}
{\cal M}_{r}(B^- \to  \chi_{0c} \pi^- \to \pi^+ \pi^- \pi^-) & = &\nonumber\\
M(B^{-} \to \chi_{0c} \pi^- ) \frac{1}{ s - m_{\chi_{0c}}^2 +  i 
\Gamma_{\chi_{0c}} m_{\chi_{0c}}}&\times &\nonumber\\
M( \chi_{0c} \to \pi^+ \pi^- ) + (s \leftrightarrow t).
\label{ares}
\end{eqnarray}
We have used the estimation  
$BR(B^{\pm} \to \chi_{c0} \pi^{\pm} )/ BR(\chi_{c0} \to \pi^+ \pi^-)  =$ 
$ 5 \times 10^{-7}$ found in  \cite{EGM}. 
Using the $\chi_{c0}$ data \cite{PDG}, we fix  the 
remaining parameters. \\

Following the analysis of \cite{DEHT}, we 
investigate the partial rate asymmetry in $B^{\pm} \to \pi^{\pm} P \bar P$. 
We are interested in the kinematical region, where the $P\bar P$ invariant 
mass is close to the mass of the $\chi_{c0}$ meson $m_{\chi_{0c}} = 3.415~GeV$. 
The partial decay width $\Gamma_p$ for 
$B^- \to  \pi^+ \pi^- \pi^-$, which contains nonresonant and resonant 
contributions,  is obtained by the 
phase space
integration from $ s_{l} = (m_{\chi_{0c}} - 2\Gamma_{\chi_{0c} })^2$ to 
$ s_{u} = (m_{\chi_{0c}} + 2\Gamma_{\chi_{0c} })^2$, where 
 $\Gamma_{\chi_{0c} } = 0.014 \pm 0.005~GeV$ is the width of $ \chi_{0c}$
\begin{equation}
\Gamma_p = 
 C \int_{s_{l}}^{s_{u}} ds \int_{t_{l}(s)}^{t_{u}(s)} dt~|{\cal M}_{nr} + 
{\cal M}_{r} |^2,  
\label{dwp}
\end{equation}
where $C = 1 /[(2 \pi m_B)^3 32]$.  
Similarly, $\Gamma_{\bar p}$ is the partial decay width for 
$B^+ \to \pi^+ \pi^- \pi^+$ containing the nonresonant and 
resonant contribution. 
 The CP-violating asymmetry 
is determined  by
\begin{equation}
|A| = |\frac{ \Gamma_p - \Gamma_{\bar p}}{\Gamma_p + \Gamma_{\bar p}}|.
\label{asym}
\end{equation}
In our calculations of the branching ratio the CHPT 
is used beyond its region of applicability and the obtained results 
should be taken with care.  
In the calculation of the partial decay width, however,  the kinematical 
region is constrained to the region around the $\chi_{c0}$ resonance and 
therefore our calculations for the partial decay 
asymmetry are more reliable.
 
For the selected set of parameters $g$, $\alpha_{1,2}$ 
we obtain
\begin{equation}
 A(B^+ \to \pi^+ \pi^- \pi^+) = 0.33 ~sin \gamma, 
\label{ab3p}
\end{equation}
\begin{equation}
A(B^+ \to K^+ K^- \pi^+)  = 0.45~ sin \gamma, 
\label{abKKp}
\end{equation}
while the asymmetry in the case of $B^+ \to \pi^0 \pi^0  \pi^+$ decay ranges 
from 
$(0.14 - 0.16)$  $sin \gamma$ 
and in the case of $B^+ \to \eta \eta  \pi^+$ decay ranges from 
$(0.10 - 0.17)$  $sin \gamma$.\\

We have  analyzed the partial rate asymmetry in $B^{\pm}\to P\bar P\pi^{\pm}$ 
decays  ($P = \pi^+$, $K^+$, $\pi^0$,  $\eta$), which signals  CP 
violation. The nonzero asymmetry results from the interference of the 
nonresonant decay amplitude and 
the resonant decay amplitude $B^\pm \to \chi_{oc} \pi^{\pm}$ followed by 
$\chi_{0c} \to P \bar P$. The asymmetry is found to be rather dependent on the 
choice of parameters  and is  $0.33~\sin\gamma$ ($0.45~\sin\gamma$) 
for $B^+\pi^+\pi^-\pi^+$ ($B^+\to \pi^+K^-K^+$), while it is  
smaller for $B^+ \to \pi^0 \pi^0  \pi^0$ and  $B^+ \to \eta \eta  \pi^+$ decays.  
The calculated partial rate asymmetries give useful guidelines for 
the experimental 
searches of the size of 
CP - violating angle $\gamma$.

\end{document}